\documentclass[espf]{aastex}
\usepackage{emulateapj5}

\slugcomment{To Appear in ApJ Letters, Revised 2005 April 12}

\shorttitle{Dust Emission around NGC~7538 IRS~1}
\shortauthors{De Buizer and Minier}

\begin{document}

\title{Investigating the Nature of the Dust Emission around Massive Protostar NGC~7538 IRS~1: Circumstellar Disk and Outflow?}

\author{James M. De Buizer}
\affil{Gemini Observatory, Casilla 603, La Serena, Chile}

\and

\author{Vincent Minier}
\affil{Service d'Astrophysique, DAPNIA/DSM/CEA Centre d'Etudes de Saclay, 91 191 Gif-sur-Yvette, France}

\begin{abstract}
We have obtained high resolution mid-infrared images of the high mass protostar NGC~7538 IRS~1 using Michelle on Gemini North and find that the circumstellar dust associated with this source is extended on both large and small scales. The large-scale mid-infrared emission is asymmetric about the peak of IRS~1, being more extended to the northwest than the southeast. The position angle of the mid-infrared emission is similar to the position angle of the linearly distributed methanol masers at this location which are thought to trace a circumstellar disk. However, this position angle is also very similar to that of the CO outflow in this region which appears to be centered on IRS~1. We suggest that the large-scale extended mid-infrared emission is coming from dust heated on the walls of the outflow cavities near the source. IRS~1 is also elongated in the mid-infrared on a smaller scale, and this elongation is near \textit{perpendicular} to the axis of the CO outflow (and the linearly distributed methanol masers). Because of its orientation with respect to the outflow and its estimated size (R$_{disk}$$\simeq$450 AU at 11.7 $\micron$), we propose that the small-scale elongation seen in the mid-infrared is a circumstellar disk that may be collimating the outflow from IRS~1.
\end{abstract}

\keywords{circumstellar matter --- infrared: ISM --- stars: formation ---ISM: individual (NGC 7538 IRS 1)}

\section{Introduction}

NGC~7538 is a very active region of high-mass star formation, which is located at 2.8 kpc (Blitz et al. 1982) in the Cas OB2 complex. In the southeastern region of this complex lies three compact and infrared-bright HII regions named IRS~1, IRS~2, and IRS~3 (Wynn-Williams, Becklin, \& Neugebauer 1974). This letter focuses on IRS~1 which is rich in many types of masers (CH$_3$OH, OH, H$_2$O, NH$_3$ and H$_2$CO; see Minier et al. 1998 and reference therein). Minier et al. (2001) found that various groups of methanol masers spread throughout this region coincide with resolved features in the radio continuum observed by Gaume et al. (1995). However, the most interesting aspect of the methanol maser emission in this region is the linearly distributed maser group (see Figure 4c) that has a well-defined velocity gradient along the spot distribution as seen in position/line-of-sight (LOS) velocity diagrams. It has been claimed that these masers exist in and trace a circumstellar disk undergoing Keplerian rotation (Minier et al. 2000; Pestalozzi et al. 2004).

Conversely, recent observations have shown that many linear methanol maser distributions are found to be parallel to outflow signatures from young massive stars, not perpendicular as would be expected if the masers were tracing disks. It has been suggested that these masers are somehow related to and tracing outflows from young stellar objects (De Buizer 2003, Moscadelli et al. 2002).

We present in this letter high spatial resolution images in the mid-infrared toward NGC~7538 IRS~1. Our goal is to see what the circumstellar dust environment can reveal about the nature of the exciting young (proto)stellar object that is embedded here. Also we wish to see if such observations can shed light onto what exactly the linearly distributed methanol masers are tracing at this location.

\section{Observations}

Initial mid-infrared images of NGC~7538 IRS~1 were obtaining on the night of 2003 December 13 at Gemini North Observatory using the facility mid-infrared imager and spectrometer Michelle. The detector in Michelle is an arsenic-doped silicon impurity band conduction (Si:As IBC) array with a format of 320$\times$240 pixels. When configured in its imaging mode Michelle has a 0$\farcs$099 pixel$^{-1}$ plate scale, which Nyquist samples the Gemini mid-infrared point-spread function (PSF). This plate scale translates to a field of view of 32$\arcsec$$\times$24$\arcsec$. Sky and telescope radiative offsets were removed using the standard chop-nod technique, with the nod throw and chop throw matched in direction and with an amplitude of 15$\arcsec$. Observations of NGC~7538 IRS~1 were taken through the \emph{Si-5} ($\lambda$$_c$=11.7$\micron$, $\Delta\lambda$=1.1$\micron$) and \emph{Qa} ($\lambda$$_c$=18.3$\micron$, $\Delta\lambda$=1.6$\micron$) filters. The exposure time through both filters was 320s on-source. Flux calibration was achieved by observing $\beta$ Peg at a similar airmass to NGC~7538 IRS~1. The assumed flux densities for $\beta$ Peg were 281.9 Jy at 11.7$\micron$ and 122.7 Jy at 18.3$\micron$. Derived flux densities for IRS~1 are 129.7$\pm$13.0 Jy at 11.7 $\micron$ and 235.1$\pm$35.3 Jy at 18.3 $\micron$. IRS~2 and IRS~3 are also apparent on the field, however the emission of IRS~2 is truncated to the north by the edge of the array. Therefore, we can only give accurate flux density measurements for IRS~3 of 11.3$\pm$1.2 Jy at 11.7 $\micron$ and 80.2$\pm$15.0 Jy at 18.3 $\micron$.

PSF observations of nearby HIP115590 were taken directly before and after the observations of NGC~7538 IRS~1. The full-width at half-maximum (FWHM) of the PSF star did not change from the first set of observations to the second, giving us an effective resolution for our observations of 0$\farcs$43 at 11.7 $\micron$ and 0$\farcs$54 at 18.3 $\micron$.

In order to better understand the spatial relationship between the mid-infrared emission and emission at other wavelengths, follow-up astrometric observations were obtained on 2004 September 7 again with Michelle at Gemini North. Repeated tests of telescopic pointing accuracy of the Gemini North telescope using Hipparcos Space Astrometry Mission stars showed the rms error of the pointing to be about 0$\farcs$25. These pointing tests were interspersed with observations of NGC~7538 IRS~1, acquired in the same way as the pointing test stars. Therefore, the absolute position of the mid-infrared emission of NGC~7538 IRS~1 is believed to be known to an accuracy of 0$\farcs$25. The coordinates of NGC~7538 IRS~1 were taken to be R.A.(J2000)=23:13:45.36 and Decl.(J2000)=+61:28:10.6.

\section{Results and Discussion}

\subsection{Mid-infrared emission characteristics}

The 18.3 $\mu$m contours of the region around IRS~1 are presented in Figure 1 overlaid on a 2MASS J, H, and K color-composite image. These mid-infrared images were centered on IRS~1, but also show IRS~3 and part of IRS~2. IRS~3 is resolved and appears very similar at 11.7 $\micron$ and at 18.3 $\micron$, having a slightly cometary-like morphology. The peak of IRS~2 is in our field, but emission is cut off north of this. IRS~2 is also resolved into an extended ($\sim$5$\arcsec$ in radius) source. Apparent in the 18.3 $\micron$ contours are sharp gouges in the southern part of the extended emission of IRS~2. This morphology is caused by the chopping of IRS~3 onto IRS~2 and array artifacts. Therefore the real morphology of the IRS~2 source is extended and likely circular.

IRS~1 in the mid-infrared shows very bright core with extended mid- and lower-level emission (Figure 2). Barely visible in the 11.7 $\micron$ image, IRS~1 has partially resolved lobes of emission lying $\sim$1.5$\arcsec$ to either side of the peak. More diffuse emission extends beyond this to the northwest. At 18.3 $\micron$ the two lobes of emission flanking the peak are much more apparent. Also more apparent is the extended emission to the northwest. The IRS~1 mid-infrared peak itself appears centrally heated from color temperature maps produced from our 11.7 and 18.3 $\micron$ images (Figure 2).

The peaks of sources IRS~1, 2, and 3 in the mid-infrared match up well to the 2MASS J, H, and K images. This shows that the hotter dust in the mid-infrared color temperature map is being heated by the star seen in the 2MASS images. The luminosity derived from the mid-infrared emission of IRS~1 alone yields a lower limit to the bolometric luminosity (using the technique from De Buizer et al. 2005) of 10$^{4}$ L$_{\odot}$, implying a spectral type earlier than B1.5. This is consistent with the lower limit to the luminosity given by Willner (1976) of 4$\times$10$^{4}$ L$_{\odot}$ and with the luminosity deduced from free-free emission of 5$\times$10$^{4}$ L$_{\odot}$ (Franco-Hernandez \& Rodriguez 2004). However, luminosities for IRS~1 are also estimated by other observations to be an order of magnitude larger when far-infrared to (sub)millimeter continuum emission is taken into account (e.g. 2$\times$10$^{5}$ L$_{\odot}$ in Werner et al. 1979). Far-infrared/sub-mm emission traces cold dust over larger extent ($\sim$0.5 pc) and the derived luminosity characterizes
a large portion of the star forming region rather than the hot dust associated with a single protostar (e.g. Minier et al. 2005). Therefore, estimates using these longer wavelength data are likely to be upper limits to the luminosity of IRS~1.

\subsection{Large-scale dust morphology: An outflow?}

Coincident to within the accuracies of our astrometry are the peak of the mid-infrared emission and the group of linearly distributed methanol masers that have been modeled well by a circumstellar disk (Pestalozzi et al. 2004). It is interesting to point out that the position angle of the linearly distributed methanol masers is 290$^{\circ}$, which is similar to the position angle of the extended mid-infrared emission of $\sim$315$^{\circ}$. Therefore is this large dusty disk-like emission seen in the mid-infrared a circumstellar accretion disk?  Given the scale of the mid-infrared emission this seems unlikely. Shepherd, Claussen, \& Kurtz (2001) argue that accretion disks around early B-type stars of $\sim$10 M$_{\odot}$ are expected to be on the order of 100s of AUs in radius. Likewise, the models of Yorke (1993,2004) show disks $\sim$400 AU in radius for similar mass stars. The main group of the linearly distributed methanol masers extends over $\sim$300~AU with a clear linear velocity gradient along the maser line in the inner $\sim$100 AU (Minier et al. 2000). This is consistent with an edge-on Keplerian disk with a radius of $\sim$500 AU using the model by Pestalozzi et al. (2004) and a 11~M$_{\odot}$ B1.5 star (rather than the 30~M$_{\odot}$ central star used in that work). However, the lobes on either side of the mid-infrared peak extend to $\sim$1.5$\arcsec$ from the peak, which is $\sim$4000 AU, an order of magnitude larger than what we would consider a typical circumstellar disk and a factor of 8 larger than the disk inferred by the model of Pestalozzi et al. (2004). Nevertheless, there are several examples in the literature (e.g. Zhang, Hunter \& Sridharan 1998) that claim detection of large disk-like tori around massive stars (usually seen in mm or molecular emission). It is possible that these types of sources are the larger, flattened, and rotating accretion envelopes that feed the central accretion disks around central massive protostars.

However, a more probable scenario for the large-scale mid-infrared morphology comes from the outflow from IRS~1. Outflows can punch a cavity through of the thick obscuring envelope surrounding an embedded young stellar object. Thermal emission, caused by the heating by the central (proto)star of the dust on the outflow cavity wall, can be reradiated and seen as diffuse and extended mid-infrared emission. An example of such a source is G35.20-0.74 (De Buizer et al. 2005).

For IRS~1 this scenario seems validated by CO outflow observations towards this region. The region of IRS~1-3 was mapped by Davis et al. (1998) and shows a CO outflow at a very similar position angle (315$^{\circ}$) to the mid-infrared dust elongation ($\sim$315$^{\circ}$). It also appears that this outflow is centered on IRS~1 (Figure 3) and that the blueshifted lobe of the CO outflow is located to the northwest. Since the outflow to the northwest is pointing towards us, we would expect to see the outflow cavity on this side of the source with less obscuration than the southeast due to the concealment of the southeastern lobe by the envelope of IRS~1. Therefore the outflow scenario also gives a natural explanation for the asymmetry seen in the mid-infrared images.

\subsection{Small-scale dust morphology: A circumstellar disk?}

An interesting feature in the dust color temperature map of Figure 2 is that the hot dust located around the peak in the temperature distribution is elongated at a position angle of 213$^{\circ}$. This is very close to perpendicular  ($\Delta$$\angle$$\sim$100$^{\circ}$) to the large-scale mid-infrared elongation that we above describe as being associated with the CO outflow.

Simple unsharp masks of the 11.7 and 18.3 $\micron$ images showed this elongation to be present in both images. Subsequent deconvolutions of the 11.7 and 18.3 $\micron$ images using an iterative maximum likelihood technique (Richardson 1972, Lucy 1974) bring out further detail, and we show these deconvolved images in Figure 4. For both the 11.7 and 18.3 $\micron$ data, after approximately 25 iterations the small-scale source elongation and proportions stabilize, and only features in the large-scale emission continue to decrease in size and break up into clumps. Therefore it appears that the deconvolution actually `resolves' the small-scale elongation of emission around IRS~1 since its size and proportions do not change with further iterations. Two-dimensional Gaussian fits to the 11.7 $\micron$ deconvolved image yields a position angle of the elongation of PA = 207$^{\circ}$, with FWHM$_{major}$ = 0.32$\arcsec$ and FWHM$_{minor}$ = 0.27$\arcsec$. A fit to the 18.3 $\micron$ deconvolved image gives PA = 207$^{\circ}$, FWHM$_{major}$ = 0.50$\arcsec$, and FWHM$_{minor}$ = 0.40$\arcsec$. The fact that both the 11.7 and 18.3 $\micron$ deconvolved images yield a similar angle of elongation and this elongation is apparent in the dust color temperature maps confirms that this inner elongation of the mid-infrared emission is real. Given the orientation of the source with respect to the CO outflow, we propose that this inner elongation is a circumstellar disk.

Using the 11.7 $\micron$ deconvolved image, the radius of the mid-infrared elongation is $\sim$450 AU given the major axis of elongation. As discussed in \S3.2, this is about the disk size inferred for a $\sim$10~M$_{\odot}$ star. The disk is larger at 18.3 $\micron$, but has the same proportions as it has at 11.7 $\micron$. The larger size is expected due to the fact that at 18.3 $\micron$ we are more sensitive to the cooler dust farther out in the disk than at 11.7 $\micron$.

\subsection{The nature of the radio continuum and methanol maser emission}

The cm radio continuum emission associated with IRS~1 has been interpreted as being the partially ionized emission from an outflow (e.g. Campbell 1984). However, the overall distribution of the radio continuum emission is extended at a position angle of $\sim$350$^{\circ}$. This is significantly different than the CO outflow (and large-scale mid-infrared extended emission) position angle. Gaume et al. (1995) resolved the radio continuum emission into bright clumps (Figure 4c) and propose another scenario in which the central star photoionizes the surrounding pre-existent knots of neutral molecular material creating dense, localized pockets of ionized gas. In this scenario, therefore, the location of the radio continuum emission is dependent upon exactly where these pre-existent clumps of material are. Gaume et al. (1995) also points out that there are regions around IRS~1 that are optically thick to cm emission, and the combined effects of optical depth and clump location may lead to the amorphous look of the radio continuum emission. Minier, Conway \& Booth (2001) showed that the methanol maser groups spread throughout this region are actually coincident with the clumps of radio continuum emission (Figure 4c). In this scenario, therefore, it is likely that these methanol maser groups are tracing these photo-ionized pre-existent clumps of molecular material.

Another possible explanation is that the radio continuum emission arises from a photoevaporated disk wind, as was modeled for IRS~1 by Lugo et al. (2004). The radio continuum would then trace the ionized gas wind from the disk surface. From the Gaussian fits of the deconvolved image at 11.7 $\micron$, a lower limit of the disk inclination angle is $i=32^{\circ}$, which means that the disk is significantly inclined from edge-on ($i=90^{\circ}$) so that surface emission may be observed. The morphology and position angle differences between the radio and mid-infrared continuum emission could be explained by gravitational instabilities in the disk and differences in angular resolution between VLA ($\sim$0.1$\arcsec$) and Gemini ($\sim$0.5$\arcsec$). Non-linear gravitational instabilities could lead to spiral shocks and clumps in the disk (e.g. Durisen et al. 2001), which could only be detected with adequate angular resolution as offered by the VLA. In this scenario, the masers arise in the spiral shocks and clumps. The methanol maser velocities vary between $-62$ and $-55$ km~s$^{-1}$ and fall in between the CO outflow velocity ranges (Davis et al. 1998). The maser velocity dispersion ($\sim$7~km~s$^{-1}$) is also consistent with the expected rotational velocity of such a disk.

\acknowledgments

Based on observations obtained at the Gemini Observatory (Program ID GN-2003B-SV-100),
which is operated by the Association of Universities for Research in Astronomy, Inc., under a cooperative agreement with the NSF on behalf of the Gemini partnership: the National Science Foundation (United
States), the Particle Physics and Astronomy Research Council (United Kingdom), the
National Research Council (Canada), CONICYT (Chile), the Australian Research Council
(Australia), CNPq (Brazil) and CONICET (Argentina). Support for JMDB was also provided by
Gemini Observatory. This publication makes use of data products from the Two Micron All Sky Survey, which is a joint project of the University of Massachusetts and the Infrared Processing and Analysis Center/California Institute of Technology, funded by the National Aeronautics and Space Administration and the National Science Foundation.

Facilities: \facility{Gemini North  (Michelle)}.

\clearpage

\begin{figure}
\epsscale{0.70}
\plotone{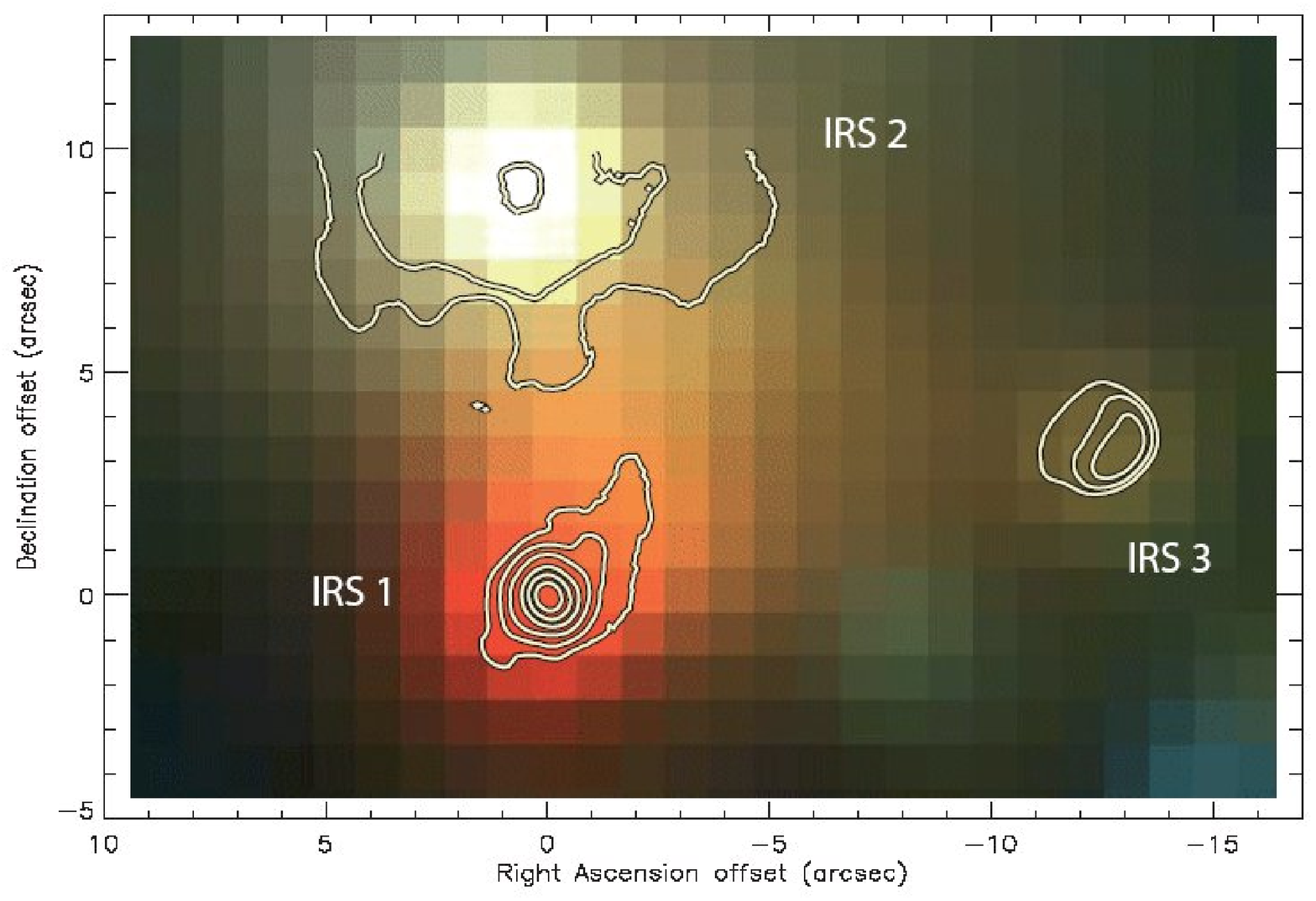}
\caption{A contour plot of the IRS~1, 2 and 3 region at 18.3 $\mu$m. These contours are overlaid on a color image that is a 3-color composite made from 2MASS J (blue), H (green), and K (red) images. The origin is the peak in the mid-infrared of IRS~1 located at R.A.(J2000)=23 13 45.36, Decl.(2000)=+61 28 10.6. \label{fig1}}
\end{figure}

\clearpage

\begin{figure}[t]
\epsscale{1.0}
\plotone{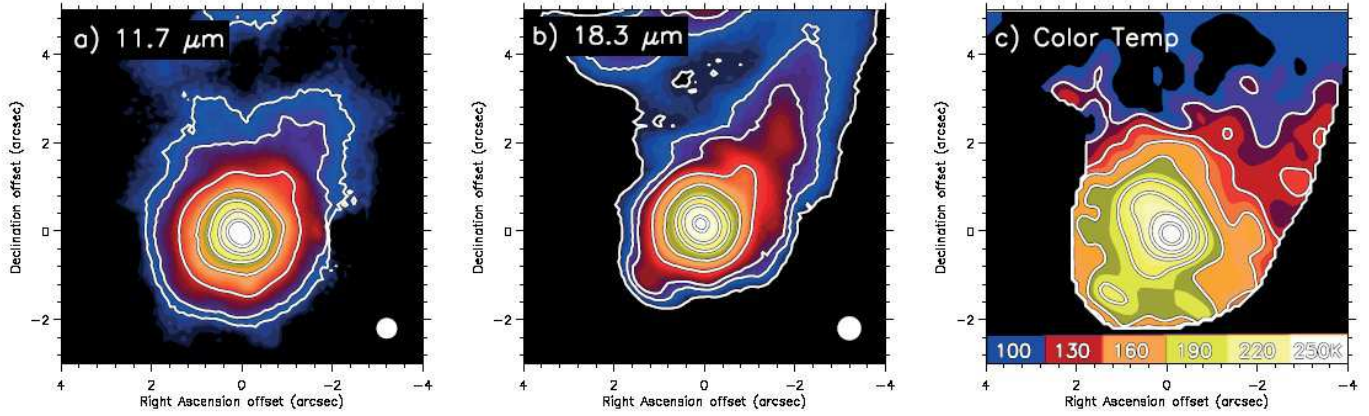}
\caption{NGC~7538 IRS~1 as seen in the mid-infrared. a) A 11.7 $\micron$ image with approximately logarithmically spaced contours at 5$\sigma$, 8$\sigma$, 22$\sigma$, 62$\sigma$, 174$\sigma$, 490$\sigma$, and 1382$\sigma$ where $\sigma$=0.68 mJy. b) A 18.3 $\micron$ image with approximately logarithmically spaced contours at 5$\sigma$, 9$\sigma$, 11$\sigma$, 24$\sigma$, 50$\sigma$, 107$\sigma$, 226$\sigma$, and 480$\sigma$ where $\sigma$=3.02 mJy. c) A dust color temperature map. The origins in each panel are the mid-infrared peak of IRS~1. The white circles in the lower right of panels a) and b) represent the size of the PSF at the given wavelength.\label{fig2}}
\end{figure}

\clearpage

\begin{figure}
\epsscale{0.50}
\plotone{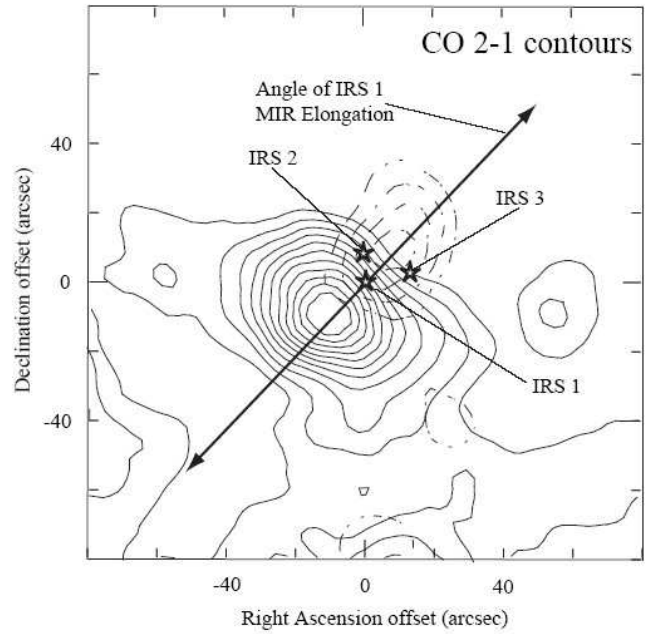}
\caption{The outflow centered on IRS 1. A line shows that the angle of elongation of the large-scale mid-infrared emission of IRS~1 is at nearly the same position angle as the CO outflow (blueshifted are dashed contours, redshifted are solid contours) from Davis et al. (1998). The positions of the mid-infrared peaks of IRS~1, 2 and 3 are marked with stars. IRS~1 also lies at the center of the CO outflow. The origin is the position of IRS~1 in the frame of the CO map. \label{fig3}}
\end{figure}

\clearpage

\begin{figure}
\epsscale{1.00}
\plotone{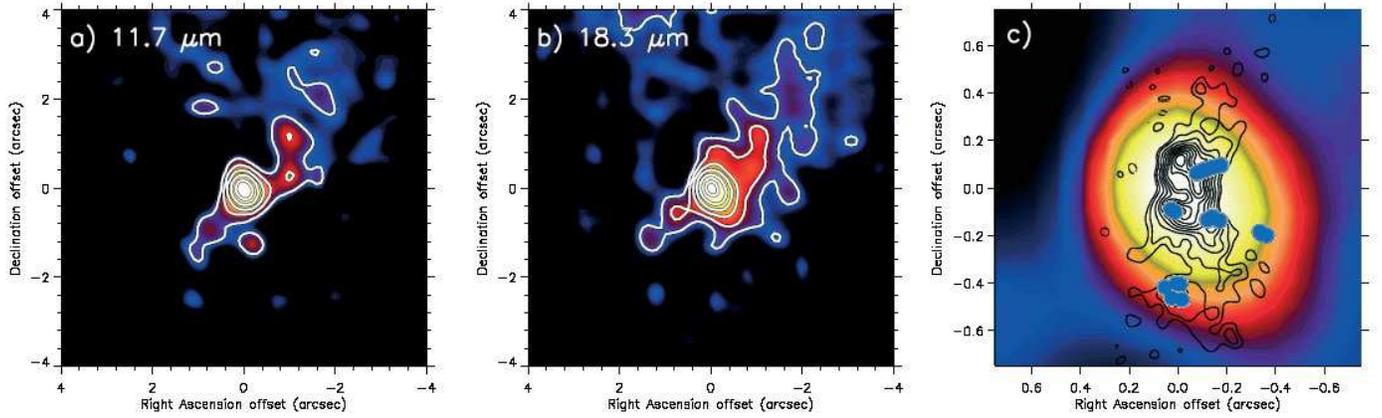}
\caption{Deconvolved images of NGC~7538 IRS~1 at a) 11.7 $\micron$ after 30 iterations, and b) 18.3 $\micron$ after 40 iterations. Panel c) is a blow-up of the 18.3 $\micron$ deconvolved image overlaid with the contours of the uniformly weighted 1.3 cm radio continuum emission of Gaume et al (1995). Also plotted (blue dots) are the methanol masers of Minier et al. (2001). The masers of interest are the linearly distributed masers to the north (seen as a blue line of dots) that have been modeled as being in a circumstellar disk. The origins in each panel are the mid-infrared peak of IRS~1. \label{fig4}}
\end{figure}

\clearpage

\end{document}